\documentclass[aps,prb,superscriptaddress,twocolumn,floatfix,showpacs]{revtex4}
\usepackage{graphicx}
\usepackage[]{amsmath}

\begin{document}
\title{Unconventional magnetic properties of the weakly ferromagnetic metal BaIrO$_3$}
\author{M. L. Brooks}
\affiliation{Clarendon Laboratory, University of Oxford, Parks Road, Oxford OX1 3PU, United Kingdom}
\author{S. J. Blundell}
\affiliation{Clarendon Laboratory, University of Oxford, Parks Road, Oxford OX1 3PU, United Kingdom}
\author{T. Lancaster}
\affiliation{Clarendon Laboratory, University of Oxford, Parks Road, Oxford OX1 3PU, United Kingdom}
\author{W. Hayes}
\affiliation{Clarendon Laboratory, University of Oxford, Parks Road, Oxford OX1 3PU, United Kingdom}
\author{F. L. Pratt}
\affiliation{ISIS Muon Facility, ISIS, Chilton, Oxon. OX11 0QX, United Kingdom}
\author{P. P. C. Frampton}
\affiliation{Inorganic Chemistry Laboratory, University of Oxford, South Parks Road, Oxford OX1 3QR, United Kingdom}
\author{P. D. Battle}
\affiliation{Inorganic Chemistry Laboratory, University of Oxford, South Parks Road, Oxford OX1 3QR, United Kingdom}

\begin{abstract} 
We present experimental evidence for small-moment magnetism below the
ferromagnetic transition temperature ($T_{\rm c3}=183$\,K) in the
quasi-one-dimensional metal BaIrO$_3$.  Further, we identify rearrangement of
the {\it local} magnetic moment distribution, which leaves the {\it bulk}
magnetization unchanged, at the Mott-like transition ($T_{\rm c1}=26$\,K).
These results are only possible via muon-spin relaxation ($\mu$SR) techniques,
since neutron scattering studies are hindered by the large absorption of
neutrons by Ir.  The low temperature characteristics of this compound, as
revealed by $\mu$SR, are unconventional, and suggest that its magnetic
properties are driven by changes occuring at the Fermi surface due to the
formation of a charge-density wave state.  
\end{abstract}


\date{\today}
\maketitle

The extended nature of the $d$ orbitals that are present in the second and
third transition series oxides means that crystalline field splittings are
enhanced and there is significant $d$--$p$ hybridization between the transition
metal ion and the surrounding oxygen octahedron. This leads to strong
coupling between the electronic, lattice, and orbital degrees of freedom, which
results in a wide variety of ground states.  Complex phase
diagrams result, and the electronic properties of these materials may be
dramatically altered by small structural changes \cite{Ohmichi, Nakamura}.
BaIrO$_3$ is a particularly interesting example: it shows weak ferromagnetism
with an unexpectedly high Curie temperature, charge-density wave (CDW)
formation, and a temperature-driven transition from a bad-metal state to an
insulating ground state \cite{Cao:CDW}.

The crystal structure of BaIrO$_3$ features three face-sharing IrO$_6$
octahedra forming Ir$_3$O$_{12}$ clusters that are vertex linked to construct
one-dimensional (1D) chains along the $c$ axis
\cite{Powell:Ferro,Egdell:BaRuO3,Siegrist:less-common}. BaIrO$_3$ is
isostructural to metallic BaRuO$_3$ \cite{Egdell:BaRuO3}, but the monoclinic
distortion in BaIrO$_3$ generates twisting and buckling of the cluster trimers
that give rise to two 1D zigzag chains along the $c$ axis and a layer of
corner sharing IrO$_6$ octahedra in the $ab$ plane bringing about both 1D and
2D structural characteristics
\cite{Powell:Ferro,Egdell:BaRuO3,Siegrist:less-common}. It is the distortions
in the structure that lead to insulating behaviour; the resistivity is further
drastically increased by substituting Ca for Ba, at the few percent level,
which introduces additional structural distortions \cite{Cao:Sr_doping}.

Magnetization measurements have demonstrated a magnetic-field insensitive
ferromagnetic transition at \mbox{$T_{\mathrm{c3}} = 175\,\mathrm{K}$}, and the
$c$-axis resistivity reveals several features: at high temperatures the
behaviour is non-metallic ($\mathrm{d}\rho_{c}(T)/\mathrm{d}T < 0$), with a
discontinuity visible at $T_{\mathrm{c3}}$; at
\mbox{$T_{\mathrm{c2}}\simeq80\,\mathrm{K}$} the resistivity peaks and the
behaviour is metallic on cooling ($\mathrm{d}\rho_{c}(T)/\mathrm{d}T > 0$)
until a Mott-like transition is encountered at \mbox{$T_{\mathrm{c1}} =
26\,\mathrm{K}$} \cite{Cao:CDW}. Non-linear conductivity and opening of an
optical gap are consistent with CDW formation accompanying the ferromagnetic
ordering.  This interpretation is supported by results of tight-binding band
structure calculations \cite{Whangbo:tight_binding} which show partially nested
pieces of Fermi surface that could signify the formation of a CDW state. The
saturation moment associated with the Ir ions, $0.03\mu_{B}$, is very small
compared to the expected moment for a $5d^5$, $S=1/2$ ion. It has been
proposed\cite{Cao:CDW,Whangbo:tight_binding} that the small moment is an
intrinsic property caused by $d$--$p$ hybridization and small exchange
splitting rather than spin canting from a localized antiferromagnetic
configuration. The addition of small amounts of Sr dopant into the material has
been shown to strongly suppress the ferromagnetic transition while increasing
the Ir saturation moment and inducing a nonmetal-metal transition at high
temperatures \cite{Cao:Sr_doping}.

Muon-spin relaxation ($\mu$SR) is an extremely sensitive probe of magnetism,
well suited to studying the spin-order and dynamics arising from the small Ir
moment; neutron scattering studies of iridates are hindered by the large
neutron absorption cross section of Ir. In order to provide a unique insight
into the magnetic properties at a local level, we performed $\mu$SR
measurements on a powdered sample of BaIrO$_3$, studying the ferromagnetic
transition at $T_{\mathrm{c3}}$ and the additional transitions \cite{Cao:CDW}
observed on cooling. The experiments were carried out using the GPS instrument
at the Swiss Muon Source, Paul Scherrer Institute, Villigen, Switzerland.  In
these $\mu$SR experiments, spin polarised positive muons ($\mu^+$, mean
lifetime $2.2\,\mu s$, momentum 28~MeV$/c$) were implanted into a powder sample
of BaIrO$_3$ prepared as described in Ref.~\onlinecite{Powell:Ferro}. The muons
stop quickly (in $<10^{-9}$~s), without significant loss of spin-polarisation.
The time evolution of the muon spin polarisation can be detected by counting
emitted decay positrons forward (f) and backward (b) of the initial muon spin
direction, due to the asymmetric nature of the muon decay \cite{musr}. In our
experiments we record the number of positrons detected by forward
($N_{\rm{f}}$) and backward ($N_{\rm{b}}$) scintillation counters as a function
of time and calculate the asymmetry function, $A(t)$, using

\begin{equation}
A(t)=\frac{N_{\rm{f}}(t)-\alpha_{\rm exp}
N_{\rm{b}}(t)}{N_{\rm{f}}(t)+\alpha_{\rm exp} N_{\rm{b}}(t)} ,
\label{asymmetry}
\end{equation}
where $\alpha_{\rm exp}$ is an experimental calibration constant and differs
from unity due to non-uniform detector efficiency. The quantity $A(t)$ is
then proportional to the average spin polarisation, $P_{z}(t)$, of muons
stopping within the sample. The muon spin precesses around a local magnetic
field, $\boldsymbol{B}_{\mu}$ (with a frequency $\nu_{\mu}=(\gamma_{\mu}/2\pi) \vert B_{\mu} \vert$, where
$\gamma_{\mu}/2\pi= 135.5~\mathrm{MHz\,T}^{-1}$).

\begin{figure}
	\includegraphics[width=8cm]{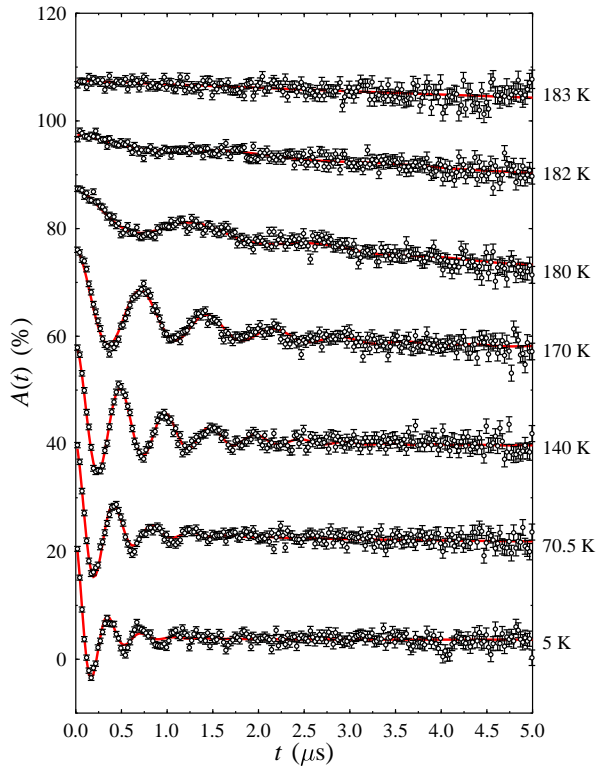}
	\caption{Typical muon asymmetry spectra measured at various temperatures. The 70~K, 140~K, 170~K, 180~K, 182~K and 183~K spectra are offset by 18~\%, 36~\%, 54~\%, 65~\%, 75~\% and 85~\% asymmetry respectively. The solid (red) lines are fits to Eq. \ref{eq:fit_func}.
	\label{fig:waterfall}}
\end{figure}

\begin{figure} \includegraphics[width=8cm]{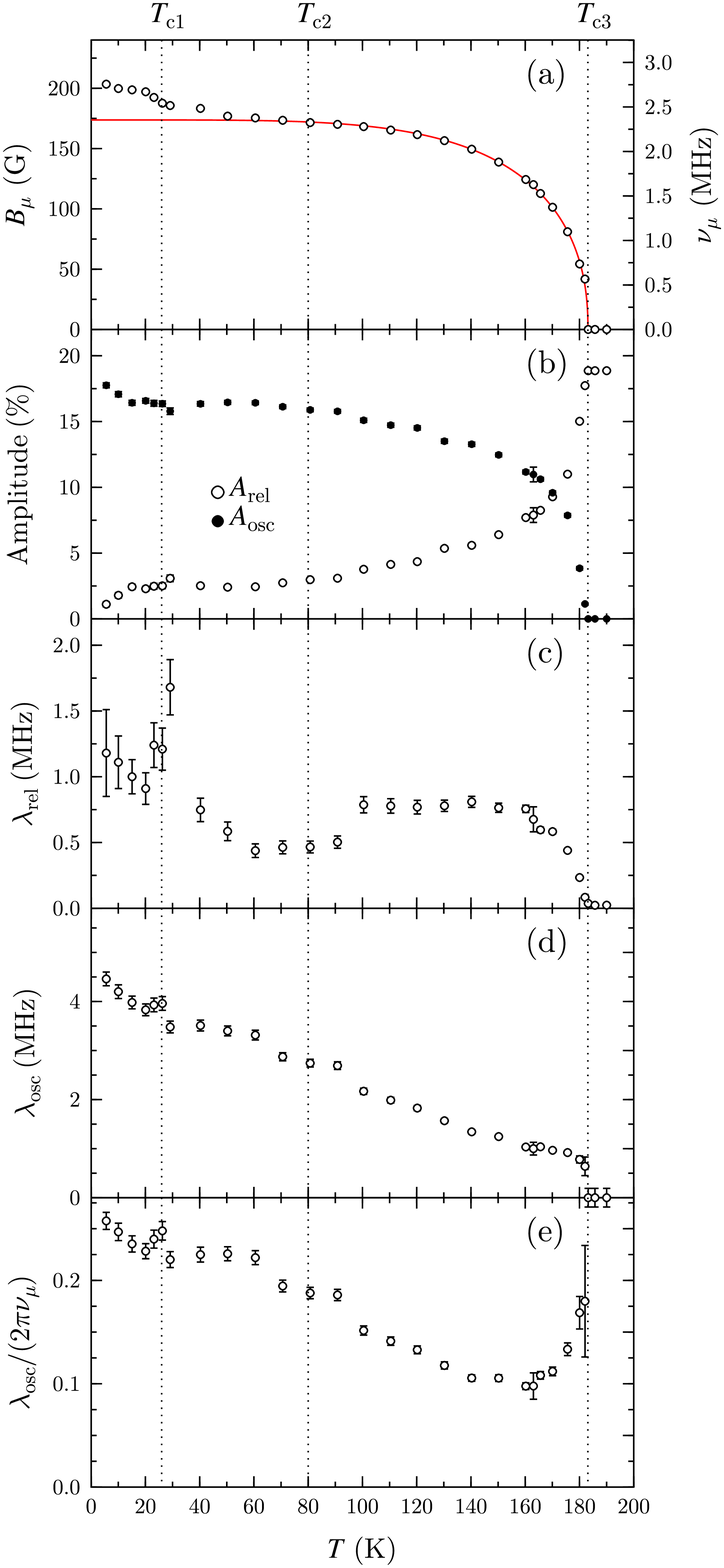}
	
        \caption{Parameters extracted from fits of data to Eq.
        \ref{eq:fit_func}. (a) Magnitude of the magnetic field at the
        muon site and a fit to Eq. \ref{eq:phenom} (solid line). (b)
        The relaxing amplitude associated with the oscillating (closed
        symbols) and the relaxing (open symbols) fractions, (c) the exponential relaxation rate ($\lambda_\mathrm{rel}$), (d) the
        oscillation damping rate ($\lambda_\mathrm{osc}$) and, (e) the ratio
        $\lambda_\mathrm{osc}/2\pi\nu_{\mu}$, which is a measure of the local magnetic
        inhomogeneity.  The dashed vertical lines mark the positions of
        $T_\mathrm{c1}$, $T_\mathrm{c2}$ and $T_\mathrm{c3}$.
        \label{fig:fitted}}

\end{figure}

Typical muon asymmetry spectra for BaIrO$_3$ are shown in
Fig.~\ref{fig:waterfall}. Clear oscillations are visible for \mbox{$T \lesssim
180\,\mathrm{K}$}, and it was possible to fit the data to the function

\begin{eqnarray}
    A(t) = A_{\mathrm{bg}} &+& A_{\mathrm{rel}}\exp(-\lambda_{\mathrm{rel}}t) \nonumber \\
    &+& A_{\mathrm{osc}}\exp(-\lambda_{\mathrm{osc}}t)\cos(\gamma B_{\mu} t)
\label{eq:fit_func}
\end{eqnarray}
over the entire temperature range studied, where $A_{\mathrm{bg}}$ represents a
time-independent background due to muons stopping in the silver foil that
surrounds the sample, $\lambda_{\mathrm{rel}}$ and $A_{\mathrm{rel}}$, are the
exponential relaxation rate and amplitude of a relaxing fraction, and
$\lambda_{\mathrm{osc}}$ and $A_{\mathrm{osc}}$ are the damping rate and
amplitude of an oscillating fraction.  The parameters extracted from these fits
are shown in Fig.~\ref{fig:fitted}.

The magnitude of the magnetic field at the muon site is shown in
Fig.~\ref{fig:fitted}(a). It begins to grow smoothly on cooling below
$T_\mathrm{c3}$, but undergoes an anomalous change near $T_\mathrm{c1}$. The
muon is able to detect changes to the local magnetisation that are not
detectable by bulk probe measurements (such as magnetization). The $B_{\mu}$
data were fitted, for $T>50\,\mathrm{K}$ (in order to avoid the anomaly), to
the phenomenological equation

\begin{equation}
	B_{\mu}(T) = B_{\mu}(0)\left(1-\left(\frac{T}{T_{\mathrm{c3}}}\right)^{\alpha}\right)^{\beta},
	\label{eq:phenom}
\end{equation}
and while the fit is good at high temperatures, there are clear departures at
low temperature (Fig.~\ref{fig:fitted}(a)). The extracted parameters are shown
in Table \ref{tab:phenom} and the fitted curve is shown in
Fig.~\ref{fig:fitted}(a) as a solid line. Our sample shows $T_{\mathrm{c3}}
\simeq 183\,\mathrm{K}$ (from both $\mu$SR and magnetization, see below), which
is a little higher than that reported previously\cite{Cao:CDW}, perhaps
reflecting the high purity of our sample.

\begin{table}
\caption{Values of parameters extracted from fitting $B_{\mu}$ to Eq. \ref{eq:phenom}. \label{tab:phenom}}
\begin{tabular}{l|r}
\hline
Parameter & Value \\
\hline
$B_{\mu}(0)\,(\mathrm{Gauss})$ & 173.7(7) \\
$T_{\mathrm{c3}}\,(\mathrm{K})$ & 183.12(4) \\
$\alpha$ & 4.5(1) \\
$\beta$ & 0.435(8) \\
\hline	
\end{tabular}
\end{table}

\begin{figure}
    \includegraphics[width=6cm]{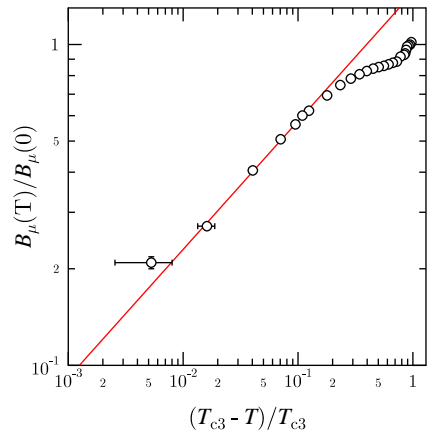}
    \caption{Plot of scaled $B_{\mu}$ against scaled temperature (using $T_{\mathrm{c3}} = 183\,\mathrm{K}$), showing a fit to points in the critical region, yielding a magnetization scaling parameter $\beta$ of 0.40(3). \label{fig:critical}}
\end{figure}

A better understanding of the critical behaviour near $T_{\mathrm{c3}}$ can be
gained from the scaling analysis, shown in Fig. \ref{fig:critical}, which
reveals that data for temperatures near to $T_{\mathrm{c3}}$ fit well to a
power law, whose gradient on the log-log plot gives $\beta = 0.40(3)$, close to 0.367 expected for three-dimensional Heisenberg behaviour. The point nearest to
$T_{\mathrm{c3}}$ is poorly fitted; this is likely due to uncertainty in our
value for $T_{\mathrm{c3}}$.

The internal field at the muon site is very much smaller (by a factor of
$\sim30$) than that measured in other magnetic oxides (see e.g.  Refs.
\onlinecite{Coldea:bilayer,Heffner:LCMO}) in which the metal ion possesses its
full moment. This is strong experimental evidence for a very small Ir magnetic
moment in BaIrO$_3$, and rules out the possibility\cite{Lindsay:cant} that the
weak ferromagnetism arises from canting of antiferromagnetically arranged
spins. Below a temperature close to $T_\mathrm{c1}$, $B_\mu$ departs sharply
from the magnetization data, showing that the Fermi surface rearrangement that
occurs at $T_\mathrm{c1}$ leads to a slight redistribution of the internal
moment distribution inside the unit cell, possibly between different Ir ions.

The amplitude of the oscillating signal $A_{\mathrm{osc}}$ is expected to be
proportional to the volume fraction of the magnetically ordered phase, and for
a conventional magnetic transition would be zero above the transition and
non-zero and constant below it (see e.g. Ref.~\onlinecite{nickelate}). In
Fig.~\ref{fig:fitted}(b), it is seen to grow slowly as the temperature is
reduced below $T_\mathrm{c3}$. Thus it is not only the size of the Ir moment,
proportional to $B_{\mu}$, which grows as the sample is cooled, but also the
fraction of the sample which is magnetically ordered. 

\begin{figure}
    \includegraphics[width=8cm]{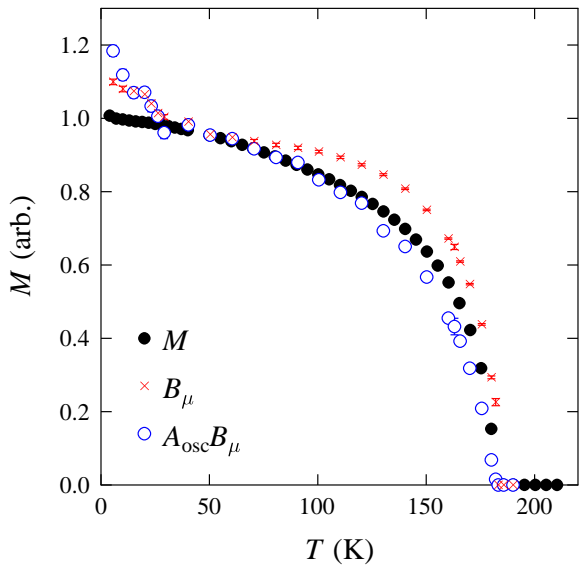}
    \caption{Scaled measurements of magnetization: field-cooled ($100\,\mathrm{Oe}$) magnetization for a powder sample of BaIrO$_3$ measured in a SQUID (filled circles), field at the muon site $B_{\mu}$ (crosses) and $B_{\mu}\times A_{\mathrm{osc}}$ (open circles). \label{fig:mag}}
\end{figure}

A comparison of the various measures of the magnetization is shown in Fig.
\ref{fig:mag}. The bulk magnetization was measured after field cooling in
$100\,\mathrm{Oe}$ in a Quantum Design MPMS SQUID magnetometer, and the
resulting (normalized) curve is seen to be significantly below the
corresponding $B_\mu$ curve over a large range of temperature. Also shown is
the result of multiplying each $B_\mu$ value by its corresponding amplitude
fraction $A_{\mathrm{osc}}$ which is expected to be proportional to the
average magnetization; this curve is a better (though not perfect) match
to the bulk measurement data. The bulk magnetization results from an ordered
fraction whose order parameter increases at the same time as its volume grows.
Since the bulk magnetization was measured in a field, it might be expected to
lie slightly above the $B_\mu \times A_{\mathrm{osc}}$ curve, which was
measured in zero applied field. We note that because the ordered fraction grows
as the sample is cooled below $T_\mathrm{c3}$, extracting a critical exponent
such as $\beta$ from either magnetization data or from the height of a neutron
Bragg peak would not be a reliable procedure for determining intrinsic
properties, whereas muons probe directly the properties of the ordered
fraction. It may be possible that the increase of the magnetic fraction with
decreasing temperature, as parameterized by $A_{\mathrm{osc}}$, reflects the
nucleation of weak ferromagnetically ordered regions between isolated
non-magnetic Ir(III) centres, since the synthesis is believed to lead to
BaIrO$_{3-\delta}$ with $\delta \sim 0.04$.

The exponential relaxation rate $\lambda_\mathrm{rel}$ reflects the dynamics of
the field at the muon site(s) and is shown in Fig.~\ref{fig:fitted}(c). A large
peak is expected near a magnetic phase transition due to the critical slowing
down of spin dynamics, but is not visible near $T_\mathrm{c3}$. It is strongly
suppressed because only a tiny fractional volume of the sample is locally
ordered at the transition temperature. A large peak is, however, seen at
$T_{\mathrm{c1}}$.  Temperature $T_{\mathrm{c2}}$ is barely evident in the
measured data: it seems to have no effect on $B_{\mu}(T)$, but possibly appears
in $\lambda_\mathrm{rel}$ as a small step near \mbox{$T = 95\,\mathrm{K}$}.

The oscillation damping rate $\lambda_\mathrm{osc}$ (shown in
Fig.~\ref{fig:fitted}(d)) is proportional to the width of the ordered field
distribution that gives rise to the oscillating signal, and is seen to rise as
the temperature is lowered below $T_\mathrm{c3}$. It has an almost linear
dependence on temperature, and seems insensitive to the transitions at
$T_\mathrm{c1}$ and $T_\mathrm{c2}$. In more conventional magnets this rate
peaks near the transition and becomes smaller on cooling as the order becomes
more uniform and static. The temperature dependence of the local magnetic
inhomogeneity, quantified by the ratio $\lambda_\mathrm{osc}/(2\pi\nu_{\mu})$
is shown in Fig.~\ref{fig:fitted}(e) and shows that, at least in the
environment of the muon site, the system becomes progressively more
magnetically inhomogeneous on cooling, in direct contrast to the behaviour observed in
most ferromagnetic systems.

In summary, we have used $\mu$SR to follow the development of magnetic order in
BaIrO$_3$ from a local viewpoint. Our experiments show that the weakly
ferromagnetic state, formed alongside the CDW state on cooling below
$T_{\mathrm{c3}}$, demonstrates unusual behaviour in the development of the
magnetically ordered volume fraction and the longitudinal and transverse
relaxation rates. This, together with the very low frequency of the muon
oscillation, leads us to conclude that the weak magnetism arises because of
small exchange splitting, and is primarily driven by the changes at the Fermi
surface that lead to the formation of the CDW state. In addition, a small
anomalous change is seen in the local magnetic field (but not the bulk
magnetization) at the Mott-like transition at $T_{\mathrm{c1}}$, and is likely
due to a local rearrangement of the magnetic moments caused by further changes
at the Fermi surface. Evidence for the metal-insulator transition at
$T_{\mathrm{c2}}$ is missing from both specific heat data\cite{Cao:spec_heat}
and our own $\mu$SR data, supporting the suggestion\cite{Cao:CDW} that
$T_{\mathrm{c2}}$ is a crossover point between partial Fermi surface gapping at
$T_{\mathrm{c3}}$ and full gapping at $T_{\mathrm{c1}}$, rather than a true
phase transition.

Parts of this work were performed at the Swiss Muon Source, Paul Scherrer
Institute, Villigen, Switzerland. We are grateful to H. Luetkens and A. Amato
(PSI) for experimental assistance, to M.-H. Whangbo (N. Carolina State Univ.)
for useful discussions concerning his tight-binding calculations and to the
EPSRC (UK) for financial support.

\end{document}